# Side-chain conformational changes upon protein-protein association

**Anatoly M. Ruvinsky[a], Tatsiana Kirys[a,b], Alexander V. Tuzikov[b], and Ilya A. Vakser[a,c,1]**

[a]Center for Bioinformatics, The University of Kansas, Lawrence, Kansas 66047

[b]United Institute of Informatics Problems, National Academy of Sciences, 220012 Minsk, Belarus

[c]Department of Molecular Biosciences, The University of Kansas, Lawrence, Kansas 66045

[1]To whom correspondence should be addressed. E-mail: vakser@ku.edu

Text pages: 21, Figures: 6

## Abstract

Conformational changes upon protein-protein association are the key element of the binding mechanism. The study presents a systematic large-scale analysis of such conformational changes in the side chains. The results indicate that short and long side chains have different propensities for the conformational changes. Long side chains with three or more dihedral angles are often subject to large conformational transition. Shorter residues with one or two dihedral angles typically undergo local conformational changes not leading to a conformational transition. The relationship between the local readjustments and the equilibrium fluctuations of a side chain around its unbound conformation is suggested. Most of the side chains undergo larger changes in the dihedral angle most distant from the backbone. The amino acids with symmetric aromatic (Phe and Tyr) and charged (Asp and Glu) groups show the opposite trend where the near-backbone dihedral angles change the most. The frequencies of the core-to-surface interface transitions of six nonpolar residues and Tyr exceed the frequencies of the opposite, surface-to-core transitions. The binding increases both polar and nonpolar interface areas. However, the increase of the nonpolar area is larger for all considered classes of protein complexes. The results suggest that the protein association perturbs the unbound interfaces to increase the hydrophobic forces. The results facilitate better understanding of the conformational changes in proteins and suggest directions for efficient conformational sampling in docking protocols.

## Introduction

Protein structure-function relationships with the focus on conformational changes upon protein-protein association have been the subject of extensive research, including systematic studies on protein sets (1-7) and specific proteins (8-13). The theory of such conformational changes has been evolving from the early "lock-and-key" concept (14), through the induced-fit model (15), to the paradigm of the conformational selection (16-20). The knowledge and understanding of these conformational changes have been accumulated and implemented in algorithms for predicting the structure of protein complexes, as evidenced by the CAPRI experiment (21). Still the conformational changes upon the formation of a complex are one of the greatest challenges for researchers studying protein interactions. A direct way to tackle this problem is to study the differences between the unbound and the bound structures of the same protein (1, 2, 6) or the differences between the alternative conformations in unbound proteins (4, 5). An encouraging factor is the growth of the PDB (22), which has been the source for studies of the side-chain conformations in proteins in general (23-30). In the 90's when only a few proteins had both bound and unbound structures known (31), Betts and Sternberg (2) studied 39 pairs of bound and unbound proteins, with only eight of the complexes having unbound structures of both binding proteins. Recently side-chain transitions were analyzed on a set of 124 protein complexes with known unbound structures (6). Currently, such sets (called docking benchmark sets because of their primary use in docking validation) contain a significantly larger and growing number of complexes (32, 33). Our DOCKGROUND non-redundant benchmark set (33) used in this study has 233 protein complexes, 99 of them having unbound structures of both binding proteins (134 complexes have unbound structure for one of the proteins).

Protein structures reveal a rich variety of conformational changes that occur at different scales upon binding (34). This includes domain motions, local folding-unfolding transitions, transitions between regular secondary structure elements in "chameleon sequences" (35, 36), disorder-to-order transitions (36, 37), and other changes in protein backbone and side chains. Although in general the different types of the conformational changes may be inter-related, in this study we focus on the conformational changes in the side chains. This choice is motivated by the fact that the majority of protein complexes in the non-redundant benchmark sets (32, 33) have small $C^{\alpha}$ RMSD between bound and unbound structures. Indeed, 71% of the DOCKGROUND set (33) used in this study has $C^{\alpha}$ RMSD < 2Å for 71% of the complexes. The Benchmark set from Weng's group (32) has interface $C^{\alpha}$ RMSD < 2.2Å for 84% of complexes. Thus studying conformational changes in the side chains is important for the development of better protein-protein docking procedures (38-42). The focus on the side chains also follows the "divide-and-conquer" paradigm: elucidating the side-chain conformational changes first, then proceeding to the backbone flexibility, and eventually to their combination (planned for our future study).

Previous studies related to the side-chain conformational changes analyzed the dynamics of the changes (43-47). The scale of the conformational change was found to be determined to a significant extent by the residue's surrounding (the environment effect). The effect appears as a decreased number of rotamers in the buried residues in comparison with the surface residues (43, 44, 46), as a small RMSD between bound and unbound states of pocket side chains (3), or as reduced fluctuations of the center of mass of such residues (48). The side-chain dynamics made it possible to differentiate the roles of the interface residues in binding, and develop a concept of anchor and latch residues that show restricted mobility and pass through similar conformations in molecular dynamics trajectories of the bound and unbound states (43, 47). The concept was extended to include conserved residues at protein interfaces with similar properties (46). The influence of the conformational changes on the binding entropy has been studied (45). Guharoy et al (6) found that the interface residues undergo more significant conformational changes and often have higher energies than the other surface residues.

Our study presents an analysis of the conformational changes of the core and the surface side chains accompanying non-covalent protein heterodimerization. We show that the mechanism and the scale of the conformational changes depend on the side chain length and the proximity of the dihedral angle to the protein backbone. Long side chains, with three or more dihedral angles, are more often subject to large conformational transitions (~$120^o$ of $\chi$ angle change). Shorter residues, with one or two dihedral angles, typically undergo small conformational changes (~$40^o$) leading to local readjustments. We suggest that the local readjustments result from the equilibrium fluctuations of the side chain around its unbound conformation. The results show that about one tenth of the complexes in our study went through the local interface changes only. All other complexes are subject to the interplay of the large conformational transitions and the local readjustments. In most residues, the largest conformational changes occur in the dihedral angle most distant from the backbone. The opposite trend is found in the residues with symmetric aromatic (Phe and Tyr) and charged (Asp and Glu) groups, where the $\chi$ angle closest to the backbone changes most. The study also reveals the interface conformational changes leading to disorder-to-order transitions and changes of the residue surface area that result in core-to-surface and surface-to-core transitions.

## Results and Discussion

Comparison of the dihedral angles values in bound and unbound residues was performed on the DOCKGROUND docking benchmark set containing the bound and unbound structures of same proteins (see Methods). The results (Fig. 1) reveal two trends: (1) generally the extent of the conformational changes increases with the increase of the number of dihedral angles in the side chain, and (2) the extent is larger for the surface interface residues than for the surface non-interface and the core residues. The relatively smaller conformational changes in the core can be explained by the tight packing. A number of the surface non-interface residues are part of the crystal packing interfaces. The relatively smaller conformational changes in the surface non-interface residues may suggest that the crystal packing interactions on average are weaker than interactions across the biological interfaces. However, the exact contribution of the crystal packing effect in the non-interface residues is beyond the scope of this study, which is focused on the analysis of the residues at the biological interfaces.

The results show that Pro, Cys and His have larger conformational changes on the non-interface surface than at the interface (Fig. 1). However, the increase is not statistically significant. The average RMSD of the interface residues with one, two, three, and four dihedral angles is 0.75, 1.22, 1.94 and 2.54Å, and the average root-square deviation of the dihedral angles (RSD, see Methods) is 40.5°, 55.1°, 111.3° and 135.0°, correspondingly. Since the dihedral angle tend to cluster near 180°, 60°, and -60° (the *trans*, *gauche*$^+$, and *gauche*$^-$ conformations), one can conclude that the side chains with one or two $\chi$ angles undergo local conformational changes, whereas the side chains with three or four $\chi$ angles can undergo a conformational transition between the energy minima. Moreover, since the average RSDs in the long side chains (the ones with more $\chi$ angles) vary around 120° (distance between two adjacent energy minima), one can assume that the conformational transitions most likely occur in a single χ angle. Other dihedral angles in the long side chains as well as those in the short side chains typically undergo a local readjustment upon the binding. It is worth noting that all the long side chains are polar, except Met. Among the side chains with two and one χ angles, Asn, His and Ser, which show the largest average dihedral angle RSD in the group, are also polar. Three non-polar residues Cys, Pro and Phe, and polar Tyr have the smallest changes of dihedral angles. Cys and Pro are the least variable in terms of RMSD. One can assume that the difference in the degree of conformational changes of polar and nonpolar residues may result from different packing around these residues. The nonpolar residues have high propensity for the tightly packed protein core, whereas the polar residues often have exposed conformations that loosen their structural surrounding allowing more space for change.

The local readjustments of the short side chains likely occur due to the thermal fluctuations of dihedral angles. Since the thermal fluctuations deviate on average ±20° from the equilibrium (49), one can estimate the average χ angle RSD (Eq. 1) due to the thermal fluctuations as $\Delta\bar{\chi}^{T} \approx 40\sqrt{n}$. For the side chains with one or two χ angles ($n$ = 1 or 2) we obtain $\Delta\bar{\chi}_{1}^{T} = 40^{o}$ and $\Delta\bar{\chi}_{2}^{T} = 56.6^{o}$ which are in excellent agreement with the statistically-derived average RSDs $\Delta\chi_{1} = 40^{o}$ and $\Delta\chi_{2} = 55.1^{o}$ (see above). The thermal fluctuations likely play a role of the “lock-and-key lubricant” providing plasticity of interfaces needed for the exact fit upon binding. The average RSDs $\Delta\chi_{3} = 111.3^{o}$ and $\Delta\chi_{4} = 135.0^{o}$ in the longer side chains deviate from the fluctuations-based estimates $\Delta\bar{\chi}_{3}^{T} = 69.3^{o}$ and $\Delta\bar{\chi}_{4}^{T} = 80^{o}$. The deviations increase with the increase of the number of the χ angles from 2 to 4. The increase of the deviations may be explained by the ability of the longer residues to establish interactions across interface earlier than the shorter residues. Indeed, binding proteins “optimize” conformations of the long residues at the early stages of their approach. The short residues get involved at later stages of binding, and thus have less time for conformational sampling.  The increase of the deviations also points toward a greater role of the induced fit mechanism in comparison to the "lubricated lock-and-key" mechanism for the longer side chains. The results show that the "lubricated lock-and-key" mechanism only is present in 11% of the complexes in the set. The rest 89% of the complexes are subject to the interplay of both the induced fit and the "lubricated fit" mechanisms. 66% of the complexes reveal 1 to 4 conformational transitions per interface, which on average consists of 18 residues.

The share of the conformational transitions, defined as conformational changes $\geq 100°$ in one of $\chi$ angles, among all the conformational changes is larger for the side chains with three or four $\chi$ than for the side chains with one or two $\chi$ (Fig. 2). The conformational transitions at the interface occur more frequently than on the non-binding surface and in the core. This observation is in agreement with the results of Guharoy et al (6) obtained on a smaller set of complexes using different definitions of surface and interface. The data in Fig. 2 show that only His residue has a higher frequency of the conformational transitions on the non-binding surface. Thr, Cys, Ile, Asn and Phe have similar frequencies of the interface and non-interface surface conformational transitions. The probability of the interface transitions $> 100°$ simultaneously in several dihedral angles decreases significantly with the increase of the number of the dihedral angles. The dihedral angle change associated with the rotation of a symmetric group in Asp, Phe, Tyr and Glu cannot exceed 90°; thus these residues do not undergo conformational transitions simultaneously in all $\chi$ angles (Fig. 2).

To further detail the picture, we computed average changes of each $\chi$ angle in the amino acids (Fig. 3). Six of the nine side chains with two dihedral angles have larger changes of the outer angle ($\chi_2$) in comparison with the near-backbone angle ($\chi_1$). The same trend is observed for all the side chains with three and four $\chi$ angles with the exception of Glu, which is slightly more prone to the changes in the first and second $\chi$ angles. Two amino-acid side chains with aromatic groups (Phe and Tyr) and two charged amino acids (Asp and Glu) demonstrate an opposite trend where the outer $\chi$ changes less than the near-backbone one. This trend is explained partly by the reduced interval of variability of the outer $\chi$ due to the symmetry of the amino acid's terminal groups.

In agreement with earlier studies (36, 37, 50) the results indicate that binding can decrease structural disorder at protein interfaces. Four percent (164 residues) of all interface residues in the set exhibited disorder-to-order transition upon binding. The disordered residues were defined as those with missing coordinates in the crystal structure. Most of the disordered residues (39%) were Ala, Gly, Glu, or Thr. On the other hand, only 11% of the disordered residues were Cys, His, Phe, Ile, Pro, Trp, or Tyr. This observation is in agreement with the classification of amino acids into order-promoting and disorder-promoting ones (37) and correlates well with the amino acids' ability to fluctuate (48).

An important conformational aspect of protein association is the changes of the residue surface area upon binding. The rate of the core-to-surface interface transitions (Fig.4), calculated as a percentage of all transitions, varied from 10-11% (Tyr, Val, and Phe) to 4% (Asn, Glu and Lys). Examples of the core-to-surface interface transitions are shown in Fig. 5 and 6. The rate of the surface-to-core interface transitions varies from 2% (Pro) to 8% (Met). Interestingly, on average, the rate of the core-to-surface transitions exceeds that of the surface-to-core transitions for all amino acids, except Asn. The largest difference between the rates is observed in six nonpolar residues, Ala, Val, Pro, Ile, Leu, Phe, and a polar Tyr. At protein interfaces, Tyr often has been identified as a hot spot (51). The bias in the core-to-surface transitions towards the nonpolar residues suggests that protein-protein interactions may perturb an unbound interface to increase the nonpolar interface area, thus increasing the hydrophobic contribution to the binding free energy (a major force stabilizing protein complexes). To test this hypothesis, we calculated average changes of polar and nonpolar interface areas induced by binding. On average, the binding increases both the polar and nonpolar interface areas in the complex. However, the increase of the nonpolar area is greater for all classes of complexes:

$\Delta S_p = 17.4 \pm 7.5 \text{Å}^2$ and $\Delta S_n = 38.3 \pm 17.2 \text{Å}^2$ for antibody/antigen;

$\Delta S_p = 25.9 \pm 7.8 \text{Å}^2$ and $\Delta S_n = 32.9 \pm 12.8 \text{Å}^2$ for enzyme/inhibitor; and

$\Delta S_p = 12.9 \pm 6.1 \text{Å}^2$ and $\Delta S_n = 24.2 \pm 9.7 \text{Å}^2$ for other ( $\Delta S_{n,p}$ is the change of the nonpolar or polar interface area).

Two typical scenarios of the core-to-surface transitions were observed, as illustrated in Fig. 5 and 6. In the first scenario, a core side chain does not change conformation, but other residues in the vicinity change conformations to increase the side-chain surface (e.g., Phe41 and Lys224 in Fig. 5 and Leu102, Pro107 and Met129 in Fig. 6). In the second scenario, both the side chain and its neighbors change their conformations (e.g., Leu99 in Fig. 5). A case where a core side chain undergoes a conformational change and its structural neighbors within 5Å stay unchanged was not observed.

## Conclusions

Knowledge of the conformational changes upon protein binding is essential for understanding molecular mechanisms of life processes and our ability to model cell phenomena. The study focuses on the side-chain conformational changes in protein heterodimerization. The results indicate that short and long side chains have propensities for different mechanisms of the conformational changes. Long side chains with three or more dihedral angles are often subject to the induced fit mechanism resulting in a conformational transition. Shorter residues with one or two dihedral angles typically undergo local conformational changes not leading to a conformational transition. The relationship between the local readjustments and the equilibrium fluctuations of a side chain around its unbound conformation is suggested. The local readjustments were dominant in 11% of the complexes. All other complexes were subject to the interplay of the induced fit and the local readjustments. We showed that most of the side chains undergo larger changes in the dihedral angle most distant from the backbone. The amino acids with symmetric aromatic (Phe and Tyr) and charged (Asp and Glu) groups show the opposite trend where the near-backbone dihedral angles change the most. The frequencies of the core-to-surface interface transitions of six nonpolar residues and Tyr exceed the frequencies of the inverse, surface-to-core transitions. The binding increases both the polar and nonpolar interface areas. However, the increase of the nonpolar area is larger for all considered classes of the protein complexes. These findings suggest that the protein association perturbs the unbound interfaces to increase the hydrophobic forces. The results facilitate better understanding of the conformational changes in proteins and suggest directions for more efficient conformational sampling in docking protocols.

## Methods

The results are obtained on a non-redundant benchmark set of 233 non-obligate protein-protein complexes from the DOCKGROUND resource http://dockground.bioinformatics.ku.edu (33, 52). The set contains unbound structures of both proteins for 99 complexes and the unbound structure of one of the proteins for 134 complexes. The structures were selected from PDB based on the following criteria: sequence identity between bound and unbound structures > 97%, sequence identity between complexes < 30%, homomultimers and crystal packing complexes excluded.

Conformational changes between unbound and bound conformations were considered for each of the protein side chains in the set. The conformational changes were expressed in terms of the root mean square deviation (RMSD) of the atoms coordinates and the root square deviation (RSD) of the dihedral angles

$$\Delta\chi = \left[ \sum_{i=1}^{n} D_i^2(\chi_i^b, \chi_i^u) \right]^{1/2}, \qquad (1)$$

where $n$ is the number of the dihedral angles in a side chain, $i$ is the index of a dihedral angle $\chi_i$, $b$ and $u$ indicate bound and unbound conformations. Function

$$D_i(\chi_i^b, \chi_i^u) = \begin{cases} |\chi_i^b - \chi_i^u|, & if\ |\chi_i^b - \chi_i^u| \le 180^o \\ 360^o - |\chi_i^b - \chi_i^u|, & if\ |\chi_i^b - \chi_i^u| > 180^o \end{cases} \qquad (2)$$

gives the shortest distance between the dihedral angels on the circle. The values of the dihedral angles were taken from 0° to 360°, except the last angles in Phe, Tyr, Asp and Glu, which were taken from 0° to 180° due to the symmetry of aromatic and charged groups (53). The dihedral angles analyzed for Arg were $\chi_{1\text{-}4}$, because the tip of the side chain containing $\chi_5$ is planar. The dihedral angles were determined using Dang program http://kinemage.biochem.duke.edu/software/dang.php. The conformational changes were placed in eighteen groups corresponding to standard amino acids (Gly and Ala were not considered). The average conformational changes and the standard deviations were computed for the interface residues (surface residues at protein interfaces), non-interface surface residues, and core residues. Surface residues were defined as those with the relative solvent accessible surface area (RASA) > 25%, as determined by NACCESS (54). Interface residues were defined as those losing > 1Å of their surface upon binding (46).

## Acknowledgments

The study was supported by R01 GM074255 grant from NIH. We thank Dr. Simon C. Lovell for the code converting dihedral angles into Cartesian coordinates of the atoms.

## Figure Legends

**Figure 1.** Average conformational change between unbound and bound residues. (*A*) Change in Cartesian coordinates (RMSD), and (*B*) change in dihedral angles (RSD, see Methods). The residues are sorted left to right according to the increasing number of χ angles, and increasing mass (if the number of χ is the same). Standard deviation is shown for the interface residues.

**Figure 2.** The share of conformational transitions. The conformational transitions are defined as those with $\Delta\chi \geq 100^o$ in any of the dihedral angles. (*A*) Percentage of conformational transitions in all conformational changes between unbound and bound structures. (*B*) Percentage of simultaneous conformational transitions in one (light gray), two (dark gray), and three (black) χ angles for interface residues. The percentage for four χ angles is negligible (not shown). The percentage of conformational changes <100° in each dihedral angle is shown by open bars.

**Figure 3.** Average dihedral angle change for different structure regions. The change between unbound and bound conformers is shown for the core (+), non-interface surface (∀) and interface (#) residues. Standard deviation is shown for the interface residues.

**Figure 4.** Frequencies of transitions between surface and core at the interface.

**Figure 5.** Core-to-surface interface transitions in porcine pancreatic trypsin induced by soybean trypsin inhibitor. The bound structure is in magenta, and the unbound one is in blue. The bound/complex structure is 1avw (55)and the unbound trypsin structure is 2a31 (56)). Phe41

keeps its conformation, while undergoing a core-to-surface transition with the relative solvent accessible surface area (RASA) change from 16.9% in the unbound state to 28.6% in the bound state, due to the conformational change in Lys60, which has two alternative unbound conformations. Lys224 keeps its conformation, while increasing its RASA from 24.1% to 28.2% due to the conformational change in Tyr217 (ΔRASA= 14.3%). Leu99 changes RASA from 19.9% to 34.4% due to its own conformational change and the change in Asn97 (ΔRASA=16.2%).

**Figure 6.** Core-to-surface interface transitions in TEM-1 β-lactamase induced by β-lactamase inhibitor protein-II. The bound structure is in magenta, and the unbound one is in blue. The bound/complex structure is 1jtd (57) and the unbound TEM-1 is 1m40 (58). Leu102 keeps its conformation, while undergoing core-to-surface transition with RASA change from 18.6% in the unbound state to 30% in the bound state, due to the conformational change in Gln99, which has two alternative unbound conformations. Pro107 keeps its conformation, while changing RASA from 22.3% to 34.1% due to the conformational changes in Tyr105 (ΔRASA=49.9%) and Lys111, which has two alternative unbound conformations. Met129 keeps its conformation, but changes RASA from 17.4% to 44.7% due to the conformational changes in Tyr105 and Lys215, which has two alternative unbound conformations. Glu104 changes RASA by 30.3%.

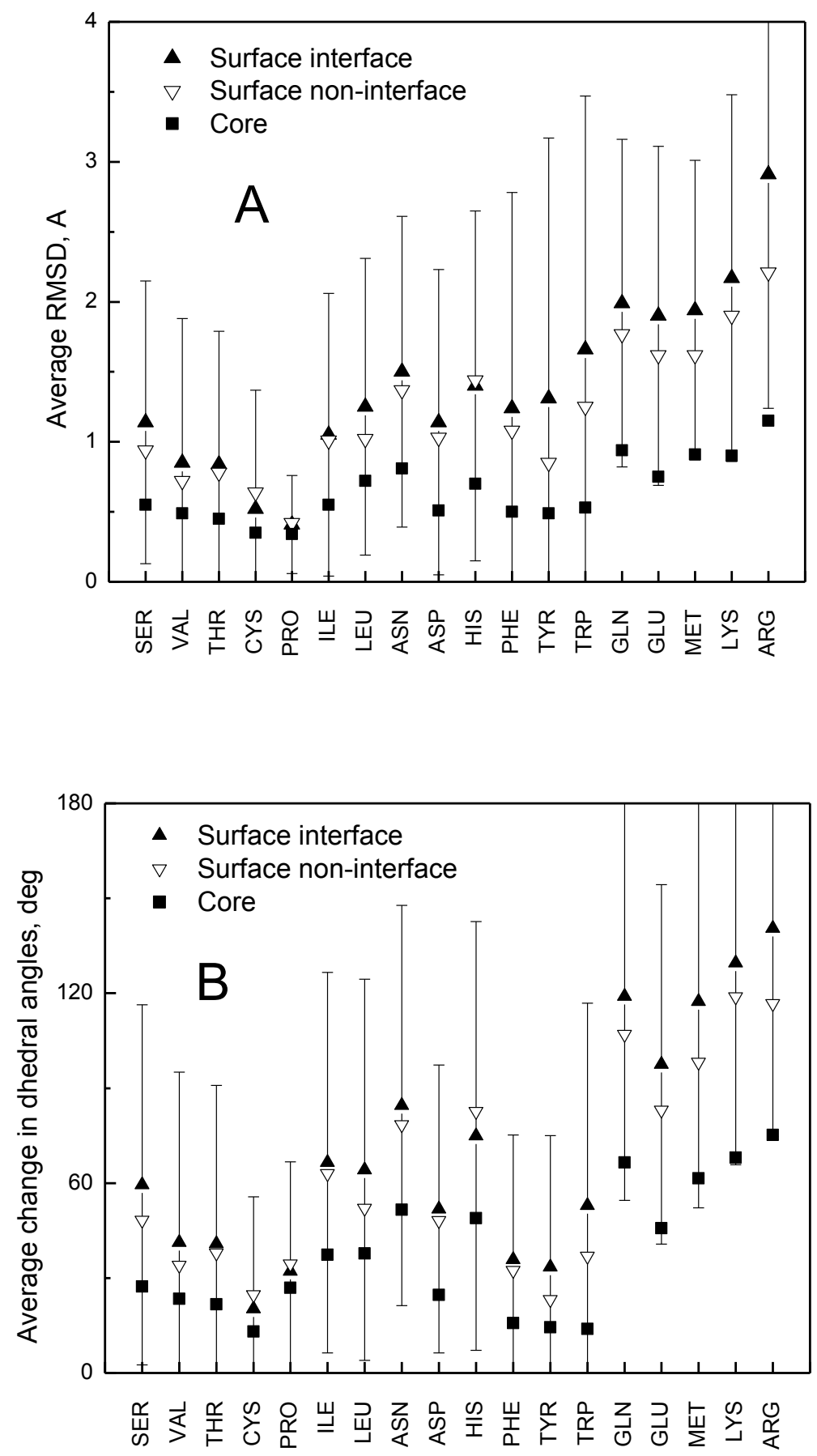

Surface interface
Surface non-interface
Core
A
Average RMSD, A
0
1
2
3
4
SER VAL THR CYS PRO ILE LEU ASN ASP HIS PHE TYR TRP GLN GLU MET LYS ARG
Surface interface
Surface non-interface
Core
B
Average change in dhedral angles, deg
0
60
120
180
SER VAL THR CYS PRO ILE LEU ASN ASP HIS PHE TYR TRP GLN GLU MET LYS ARG

Figure 1

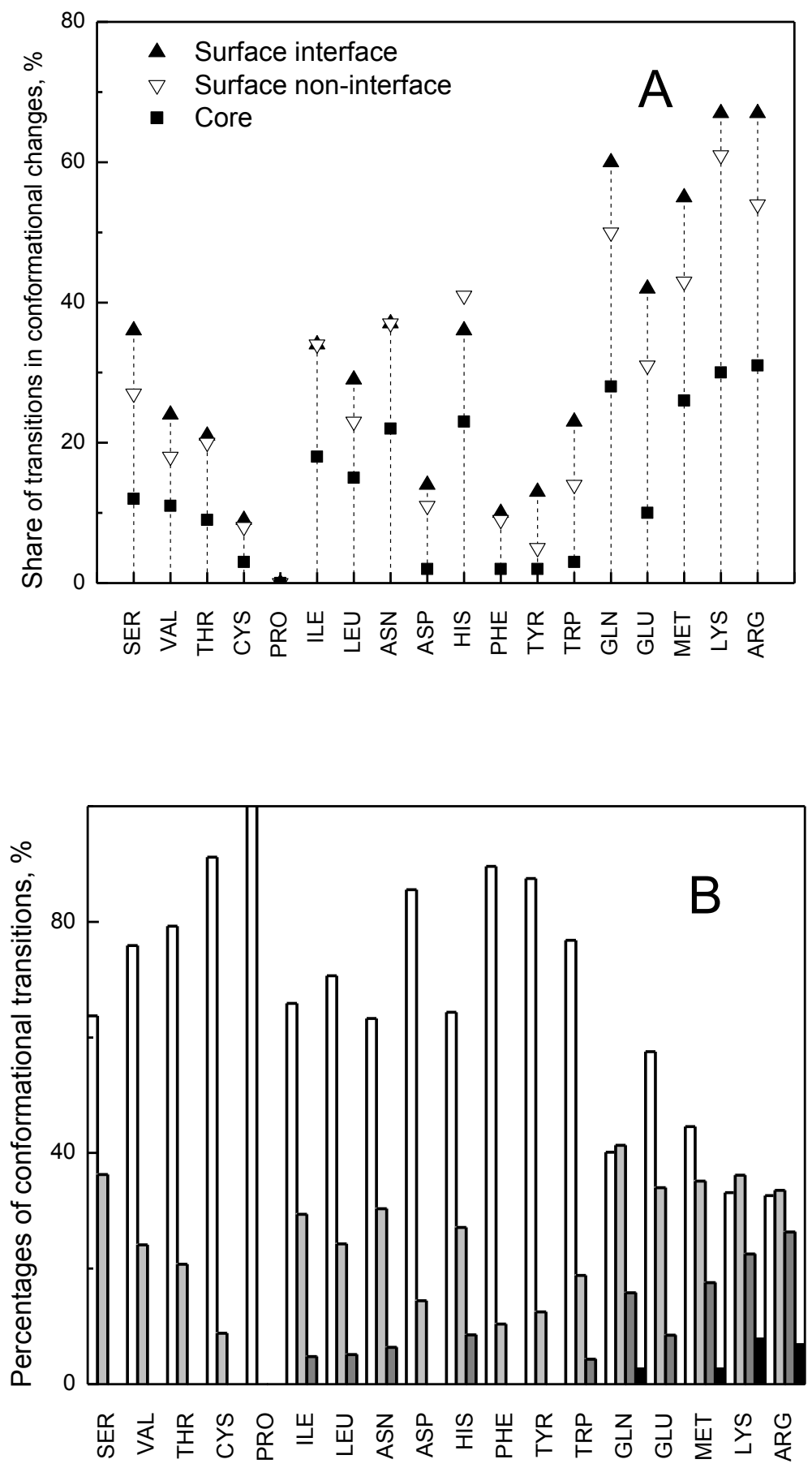
A
Surface interface
Surface non-interface
Core
Share of transitions in conformational changes, %
0
20
40
60
80
SER
VAL
THR
CYS
PRO
ILE
LEU
ASN
ASP
HIS
PHE
TYR
TRP
GLN
GLU
MET
LYS
ARG
B
Percentages of conformational transitions, %
0
40
80
SER
VAL
THR
CYS
PRO
ILE
LEU
ASN
ASP
HIS
PHE
TYR
TRP
GLN
GLU
MET
LYS
ARG

Figure 2

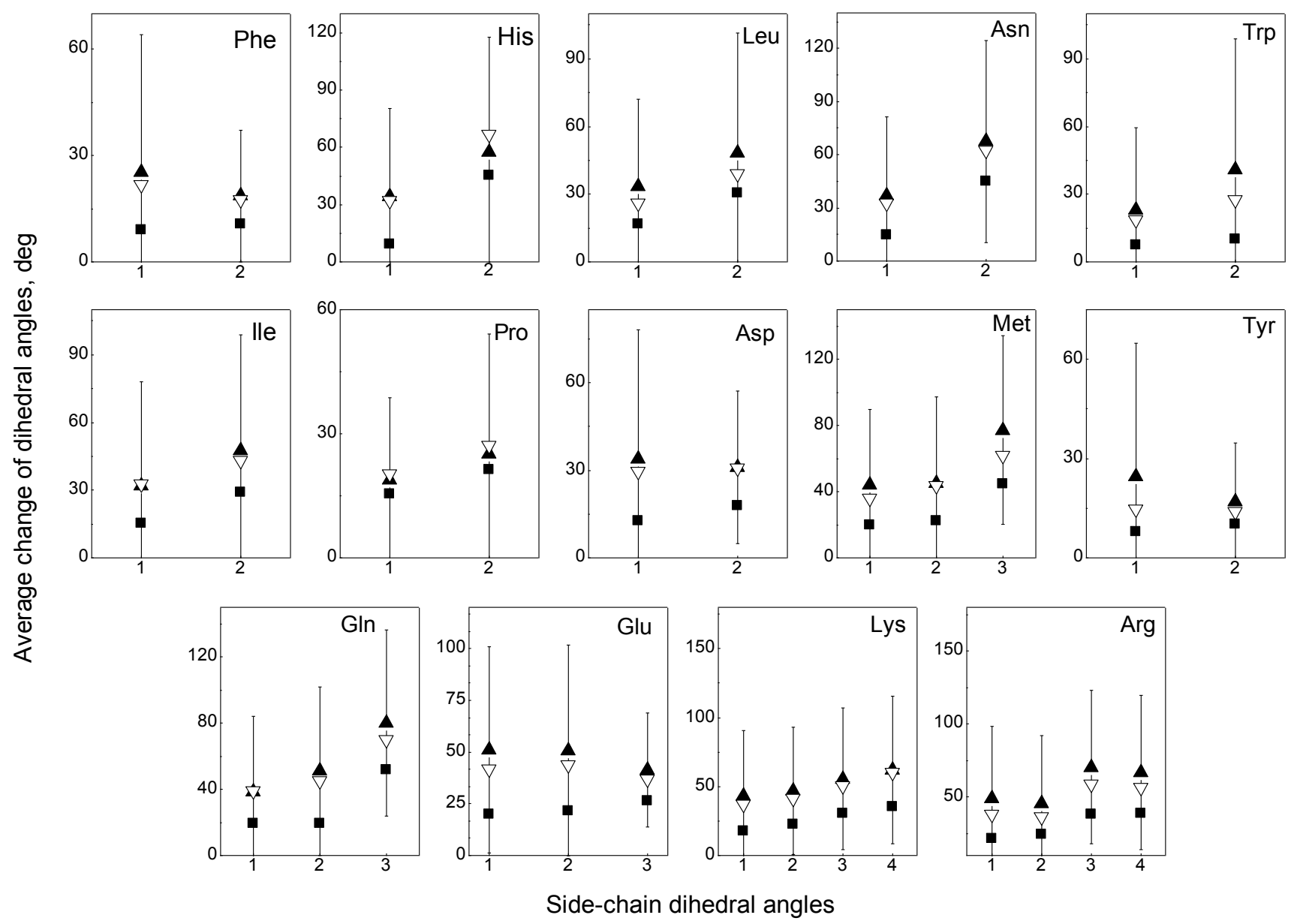
Phe
His
Leu
Asn
Trp
Ile
Pro
Asp
Met
Tyr
Gln
Glu
Lys
Arg
Average change of dihedral angles, deg
Side-chain dihedral angles

Figure 3

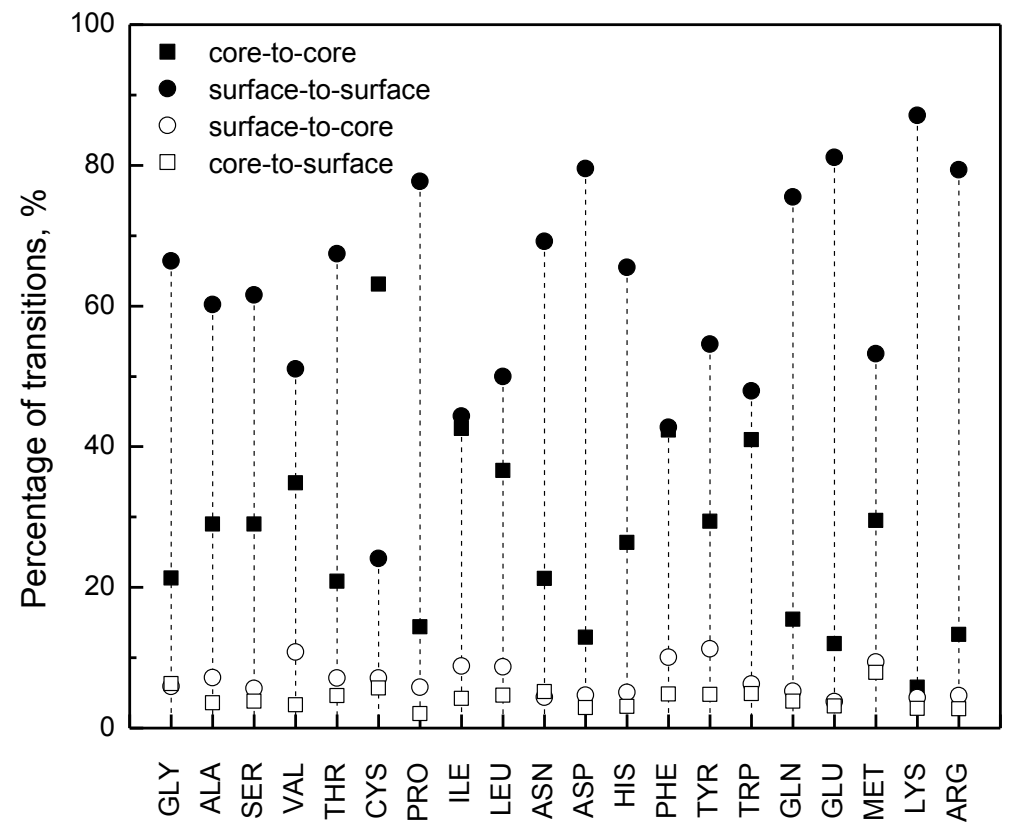
core-to-core
surface-to-surface
surface-to-core
core-to-surface
Percentage of transitions, %
0
20
40
60
80
100
GLY
ALA
SER
VAL
THR
CYS
PRO
ILE
LEU
ASN
ASP
HIS
PHE
TYR
TRP
GLN
GLU
MET
LYS
ARG

Figure 4

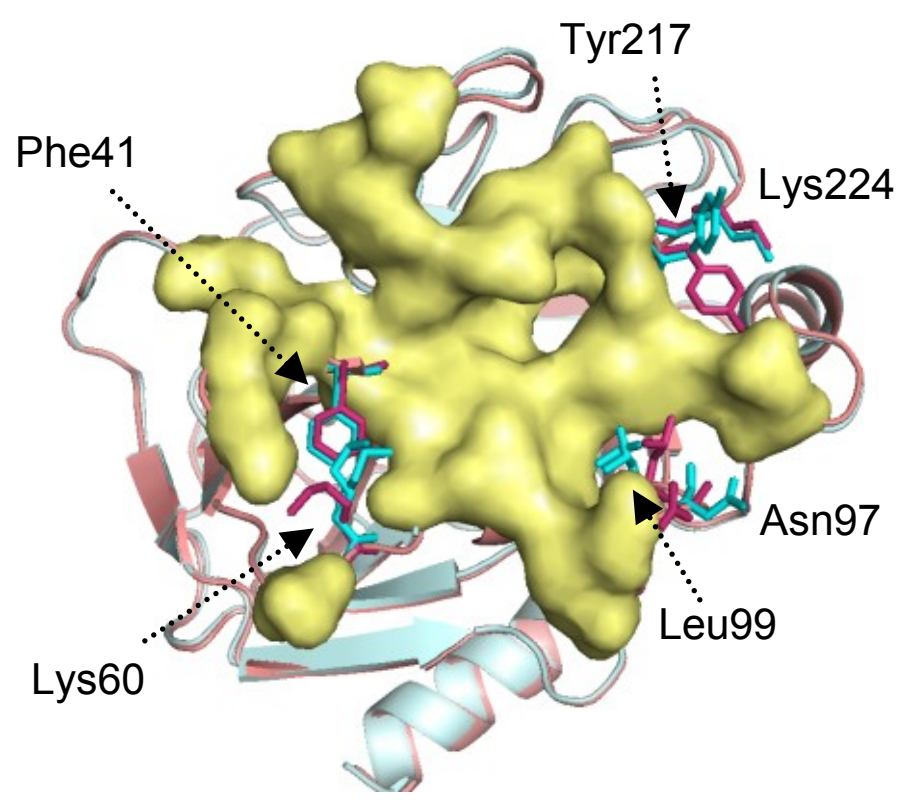

Figure 5

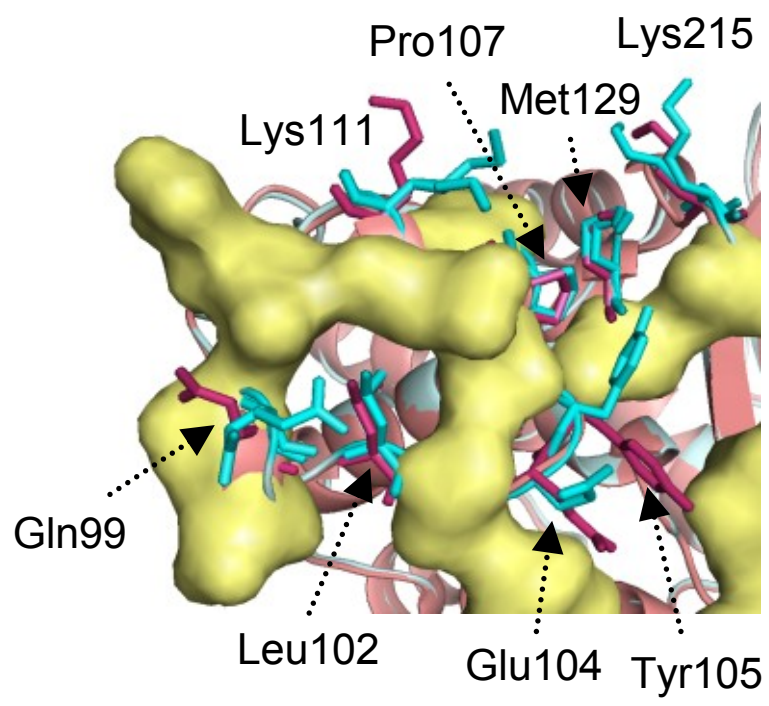

Figure 6